\documentclass[doublespacing]{elsart}

 \usepackage{graphicx}

\usepackage{amssymb}

\begin{document}

\begin{frontmatter}



\title{Development of an n-channel CCD, CCD-NeXT1, 
  for Soft X-ray Imager onboard the NeXT satellite}


\author[label0,label1]{Shin-ichiro Takagi\corauthref{cor1}},
\author[label1]{Takeshi Go Tsuru\corauthref{cor2}},
\author[label1]{Tatsuya Inui},
\author[label1]{Hironori Matsumoto},
\author[label1]{Katsuji Koyama},
\author[label2]{Hideki Ozawa}
\author[label2]{Masakuni Tohiguchi},
\author[label2]{Daisuke Matsuura},
\author[label2]{Emi Miyata},
\author[label2]{Hiroshi Tsunemi},
\author[label0]{Kazuhisa Miyaguchi},
\author[label0]{Kentaro Maeda},
\author[label0]{Hirohiko Kohno}

\address[label0]{Hamamatsu Photonics K.K., 1126-1, Ichino-cho, Hamamatsu, 
  435-8558, Japan}
\address[label1]{Department of Physics, Graduate School of Science, Kyoto
  University, Kitashirakawa, Sakyo-ku, Kyoto, 606-8502, Japan}
\address[label2]{Department of Earth and Space Science, 
  Graduate School of Science, Osaka University, Machikaneyama-cho, 
  Toyonaka-shi, Osaka, 560-0043, Japan}

\corauth[cor1]{Corresponding author.\\
 {\it Tel}:~+81-53-434-3311,\\
 {\it Fax}:~+81-53-431-0228,\\
 {\it E-mail address}:~s-takagi@ssd.hpk.co.jp~(S. Takagi)}
\corauth[cor2]{Corresponding author.\\
 {\it Tel}:~+81-75-753-3868,\\
 {\it Fax}:~+81-75-753-3799,\\
 {\it E-mail address}:~tsuru@cr.scphys.kyoto-u.ac.jp~(T.G. Tsuru)}
\begin{abstract}

{\it NeXT} (New X-ray Telescope) is the next Japanese X-ray
astronomical satellite mission after the {\it Suzaku} satellite.
{\it NeXT} aims to perform wide band imaging spectroscopy. Due
to the successful development of a multilayer coated mirror, called
a supermirror, {\it NeXT} can focus X-rays in the energy range from
0.1~keV up to 80~keV.
To cover this wide energy range, 
we are in the process of 
developing a hybrid X-ray camera, Wideband X-ray Imager (WXI) as a
focal plane detector of the supermirror. The WXI consists of X-ray
CCDs (SXI) and CdTe pixelized detectors (HXI), which cover the lower
and higher X-ray energy bands of 0.1--80~keV, respectively. The X-ray
CCDs of the SXI are stacked above the CdTe pixelized detectors of the
HXI.
The X-ray CCDs of the SXI detect soft X-rays below $\sim 10$~keV and
allow hard X-rays pass into the CdTe detectors of the HXI without loss.
Thus, we have been developing a
``back-supportless CCD'' with a thick depletion layer, a thinned
silicon wafer, and a back-supportless structure. In this paper, we
report the development and performances of an evaluation model of CCD
for the SXI, ``CCD-NeXT1''. 
We successfully fabricated two types of
CCD-NeXT1, unthinned CCDs with 625-$\mu$m thick wafer and
150-$\mu \rm m$ thick thinned CCDs. By omitting the
polishing process when making the thinned CCDs, we confirmed
that the polishing process does not impact the X-ray performance. 
In addition, we did not find significant differences 
in the X-ray performance between the two types 
of CCDs. The energy resolution and readout noise are $\sim 140$~eV (FWHM) 
at 5.9~keV and $\sim 5$ electrons (RMS), respectively. The estimated 
thickness of the depletion layer is  $\sim 80 ~\mu \rm m$.
The performances almost satisfy the requirements of the baseline plan
of the SXI.

\end{abstract}

\begin{keyword}
NeXT satellite; Soft X-ray Imager; X-ray CCD; back-supportless CCD; CCD-NeXT1
\PACS 95.55.-n, 95.55.Aq, 95.55.Ka, 
\end{keyword}
\end{frontmatter}


\section{Introduction}
{\it NeXT} (New X-ray Telescope) is the sixth Japanese X-ray
astronomical satellite mission, which is proposed to be launched
around 2012. A Hard X-ray Telescope (HXT), supermirror onboard {\it
  NeXT} (see Tawara {\it et~al.} (2003) \cite{tawara03} and references
therein), has a large collecting area for X-rays in the energy from
0.1 to 80~keV. In particular, the HXT has a high reflectivity even
in the hard X-ray band above 10~keV. 
Previous satellite missions have not had X-ray focusing optics
capable of observations in this band.
{\it NeXT} is designed to be the first to
perform imaging and spectroscopic observations in the energy band
above 10~keV. In order to meet the energy range
covered by the HXT, we have been developing a Wideband X-ray Imager
(WXI).

The first successful space flight use of X-ray CCDs as photon counting
and spectroscopic imagers was the SIS aboard $ASCA$
\cite{byrke91}. Since then, X-ray CCDs have become standard focal plane
detectors for X-ray telescopes in the X-ray energy band of
0.1--10~keV, and have been adopted as the principal detectors of recent
X-ray observatories such as the ACIS of {\it Chandra}
\cite{garmire03}, the EPIC of {\it XMM-Newton} 
\cite{struder01,turner01}, and the XIS of {\it Suzaku} \cite{koyama07} 
because X-ray CCDs have well balanced spectroscopic,
imaging, and time resolution performances. However, to achieve a quantum
efficiency of 10\% for X-rays with an energy of 40~keV, 
a $\sim 1000~{\rm \mu m}$ depletion
layer is required, which is nearly
impossible. A detector with a high-Z material is essential 
for observations in the hard X-ray band above 10--20~keV. 
On the other hand, the
performances of imaging and spectroscopy below 10~keV of
high-Z solid detectors such as CdTe detectors are poorer than those of
X-ray CCDs, suggesting that a single detector cannot cover the entire
0.1--80~keV band with the best X-ray performance. Thus, we have been
developing a hybrid camera, the Wide band X-ray Imager (WXI), 
which combines an X-ray CCD and a CdTe pixelized detector
\cite{Takahashi1999,tsuru01,tsuru04,takahashi04}. 
Holland (2003) \cite{adholland03} has also reported the first 
laboratory demonstration of such a hybrid detector 
with a thinned X-ray CCD, which is operated in front of CZT detector.

The WXI consists of two sub-instruments; the Soft X-ray Imager (SXI)
and the Hard X-ray Imager (HXI); overviews can be found in
Tsuru {\it et~al.} (2005) \cite{tsuru05} and Takahashi {\it et~al.}
(2004) \cite{takahashi04s}, respectively. The SXI and HXI are the
upper and lower parts of the WXI, respectively. The SXI consists of
X-ray CCDs with a thick depletion layer for the lower energy band
below 10--20~keV. The HXI is based on CdTe pixelized detectors, which cover
the hard X-rays above 10--20~keV.

We have developed a new type of CCD for the SXI, a ``Back-Supportless CCD''
(BS-CCD), in which the back supporting package under the imaging area
of the CCD is removed. Most X-rays absorbed in the field-free region
of the CCD are undetected and lost. Hence, we also removed the
field-free region as much as possible. The BS-CCD of the SXI is
placed over the CdTe pixelized detectors of the HXI. Soft X-rays
are detected in the BS-CCD, while hard X-rays penetrate through the
BS-CCD and are detected by the CdTe pixelized detectors. Thus, both
soft and hard X-rays are detected without loss.



As previously reported by Tsuru {\it et~al.} (2005) \cite{tsuru05} in
detail, we have been developing a BS-CCD for the SXI following two
plans of a rather conservative ``baseline plan'' and an innovative
``goal plan'' in parallel. Table~\ref{tab:SXIgoala} shows the
specifications of the BS-CCD of the two plans along with those of
the XIS \cite{koyama07}, which is one of the most excellent X-ray CCDs
currently in orbit.

In the goal plan, we realize a fully-depleted back-illuminated type
of BS-CCD with a very thick depletion layer of $\sim 200~{\rm \mu
  m}$ by adopting a p-channel device. We have already successfully
fabricated test devices with a  full depletion layer, which was 
$\sim$ 200-$\mu{\rm m}$ thick. The details and status of the
developments are reported elsewhere
\cite{kamata04,takagi05,kamata06,matuura06,takagi06,tsuru06}.


In the baseline plan, we developed BS-CCDs based on a natural extension of
our successful developments of CCD-CREST/CREST2 and
MAXI-CCD in order to minimize the risk involved with 
their development \cite{bamba01,tsunemi05,tomida00,miyata02}. 
We adopted a front-illuminated type of CCD mainly
due to the manufacturing process \cite{takagi05}. The thickness of
the depletion layer is designed to be 70--80$~{\rm \mu m}$ or more. We
have already successfully developed a small test model of BS-CCD and
confirmed that the thinning processes of
the wafer and the back-supportless structure do not degrade 
the performance \cite{takagi05}.
After the successfully developing the small test model, we constructed an
evaluation model, ``CCD-NeXT1''. Following CCD-NeXT1, we will develop
a flight model, ``CCD-NeXT2'', which matches the specifications of the
SXI shown in Table~1. In this paper, we report the development and the
performance of CCD-NeXT1\footnote{Note that  part of the results
  reported herein have already been reported as a contributing paper of a SPIE
  conference \cite{OzawaSPIE06_CCD-NeXT1}.}.




\section{Specification of CCD-NeXT1}

Table~\ref{tab:NeXT1spec} shows the specifications of CCD-NeXT1. 
CCD-NeXT1 is a frame transfer type of CCD 
with an imaging area (IA) and a frame-stored area (FS). The format
and pixel size are $2000 \times 2000$ pixels and $\rm 12 ~\mu m \times
12 ~\mu m$ both in the IA and FS, respectively. The size of IA is $\rm 24~mm \times
24~mm$. CCD-NeXT1 has two readout nodes. 

\begin{figure}
 \begin{center}
   \includegraphics[width=14cm]{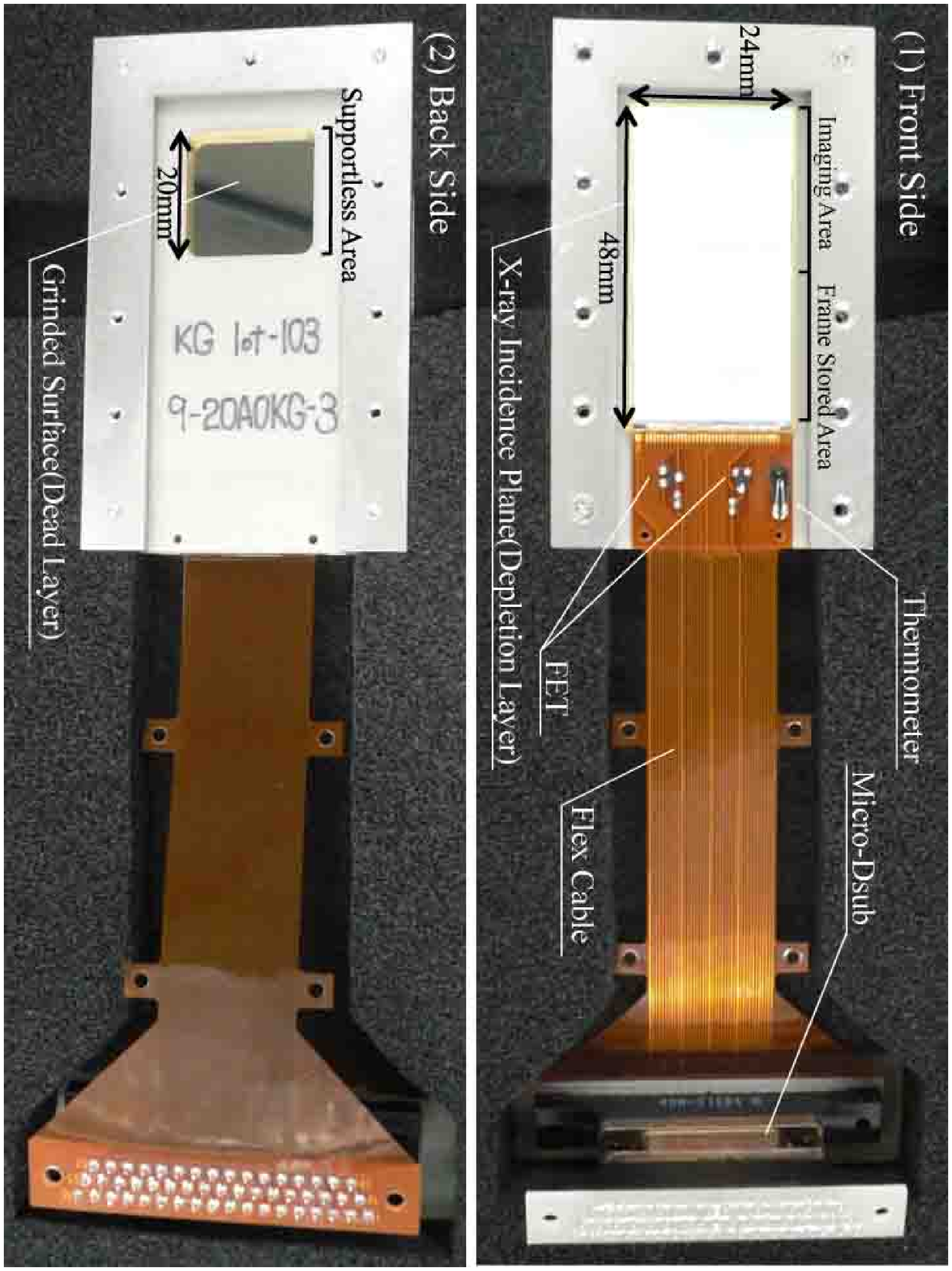}
   \caption[Picture of CCD-NeXT1]{Pictures of a CCD-NeXT1 device. (1)
     Front. (2) Back. Due to the aluminum coating for
     optical blocking, reflections on the surface of both sides are
     clearly seen.}
   \label{fig:NeXT1}
 \end{center}
\end{figure}

Figure~\ref{fig:NeXT1} shows pictures of the
CCD-NeXT1 device.
The front and back sides
of the package are open and the surfaces of both sides
of the device can be seen. Because 
the surface of the CCD-NeXT1 device is coated
with evaporated aluminum for optical blocking, the reflections on the
surface of both sides are observed in Figure~\ref{fig:NeXT1}.



First, we tested the unthinned type of CCD-NeXT1 with 
625-${\rm \mu m}$ thick silicon wafers to determine 
if the CCDs were successfully processed.
Next, we fabricated CCD-NeXT1 devices processed on
150-${\rm \mu m}$ thinned wafers and test them.
Both devices were fabricated on wafers with a high specific resistance
of $5~{\rm k \Omega \cdot cm}$.

\section{Performances of CCD-NeXT1}

\subsection{Unthinned device}

\begin{figure}
 \begin{center}
   \includegraphics[width=6.8cm]{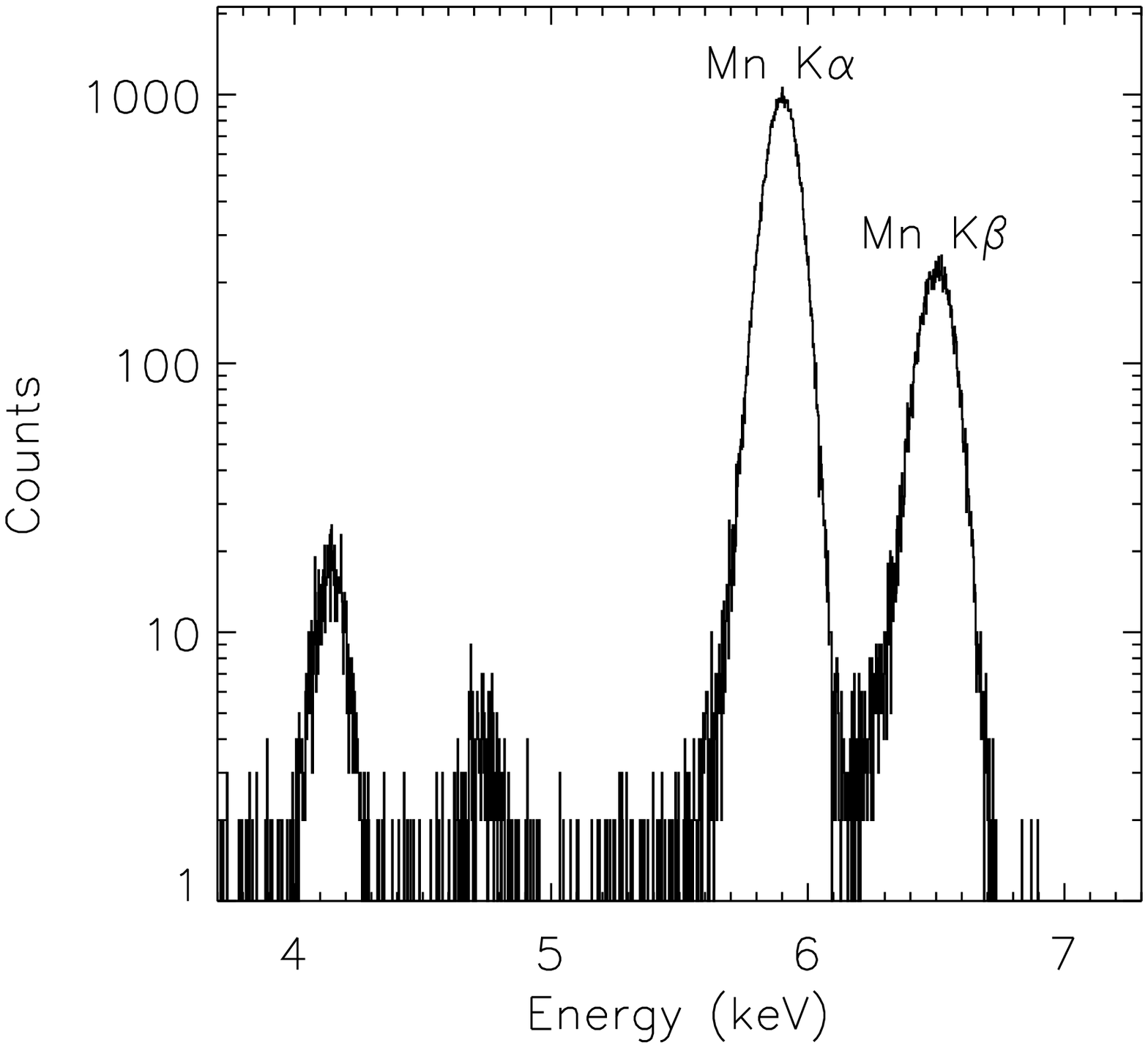}
   \includegraphics[width=6.8cm]{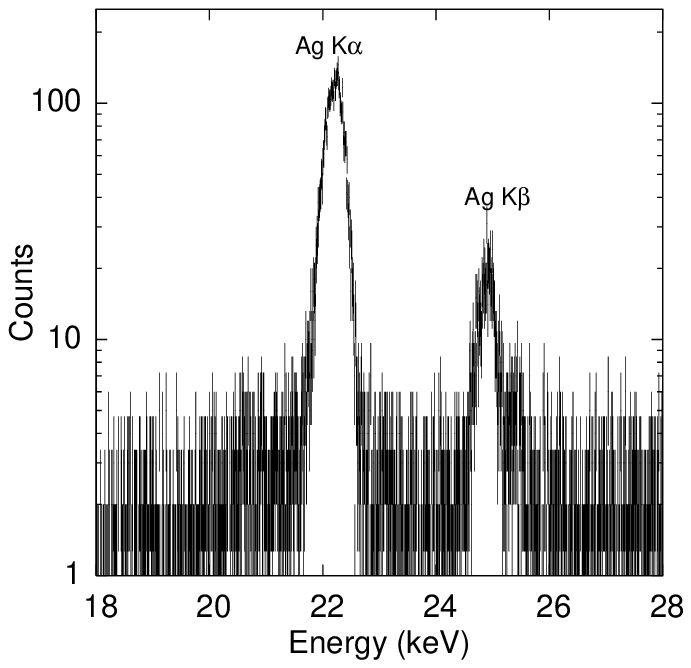}
   \caption[]{X-ray spectra of $^{55}$Fe (left) and $^{109}$Cd
(right) obtained with an unthinned device.}
  \label{fig:KG5spec}
 \end{center}
\end{figure}

We uniformly irradiated the IA of the unthinned CCD-NeXT1 device 
with $^{55}$Fe and $^{109}$Cd X-rays.
The FS was covered by a thick
aluminum plate to block incident X-rays. The CCD-NeXT1 device was
driven in a frame transfer mode and the signal was read out through
both of the nodes (hereafter referred to as nodes A and B). We performed the
clocking and readout of the device with the {\it MiKE} CCD operation
system, which was developed at Osaka University \cite{miyata02}.
Table~\ref{tab:NeXT1results} summarizes the results, while
Figure~\ref{fig:KG5spec} shows the acquired spectra from the 
$^{55}$Fe and $^{109}$Cd sources, where single pixel events and  
multi-pixel events within $3 \times 3$ pixels ($ASCA$ grade 0--7
\cite{yamashita99}) were accumulated for $^{55}$Fe and $^{109}$Cd,
respectively.
The energy resolutions of nodes
A and B for 5.9~keV X-rays were $141.8\pm 0.6~{\rm eV}$ and $146.7\pm
0.8~{\rm eV}$, respectively. We fitted a single Gaussian function to
the histograms of the horizontal over clock regions and obtained
readout noises of $4.7\pm 0.2~{\rm e^{-}}$ and $5.5\pm 0.2~{\rm
  e^{-}}$ for nodes A and B, respectively. The node sensitivity of this
device was $3.2~{\rm \mu V/e^-}$, which is higher than that of the
test model of BS-CCD \cite{takagi05}. The readout noise of CCD-NeXT1
was also the same as that of the test model of BS-CCD \cite{takagi05}.
Both vertical and horizontal CTIs are as good as those of
the test model of BS-CCD \cite{takagi05} and satisfy the requirements
for the baseline plan of SXI.
  
We measured the detection efficiency and estimated the thickness of
the depletion layer using X-rays from $^{109}$Cd. We
collimated the X-rays to expose only
the IA of the CCD-NeXT1 device. The absolute X-ray flux of the source
was well-calibrated. In this experiment, the sum of the absolute flux
of $\rm Ag~K_{\alpha}$ and $\rm Ag~K_{\beta}$ was calibrated, but the
individual contributions were not separately calibrated.
The detection rate of $\rm Ag~K_{\alpha} + \rm Ag~K_{\beta}$ was
$0.14~{\rm count~s^{-1}}$, which corresponds to a detection efficiency of
$\sim 5.2$\% and is consistent with an 84-${\rm \mu m}$ thick depletion layer.

\subsection{Thinned device}

\begin{figure}
 \begin{center}
   \includegraphics[width=10cm]{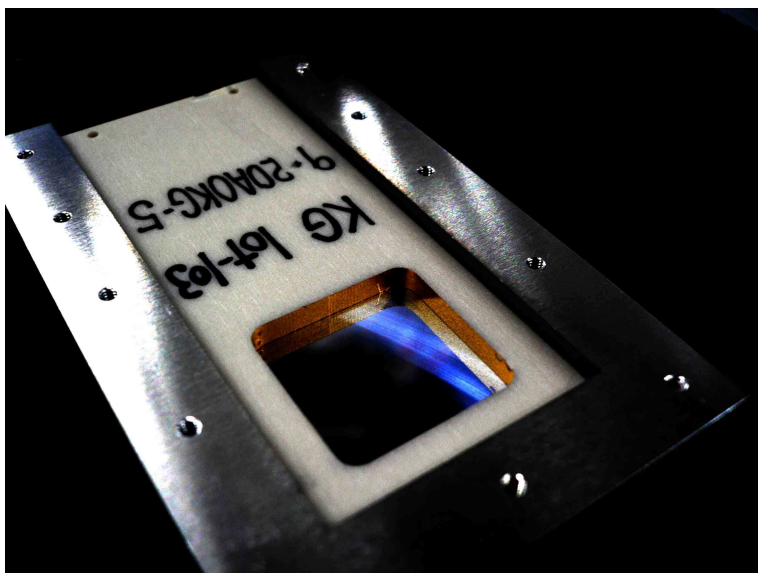}
   \caption[Tiny trails found on the grinding surface of CCD-NeXT1.]{
     Tiny trails found on the grinding surface of the thinned device of
     CCD-NeXT1.} 
  \label{fig:grind_trail}
 \end{center}
\end{figure}

\begin{figure}
 \begin{center}
   \includegraphics[width=14cm]{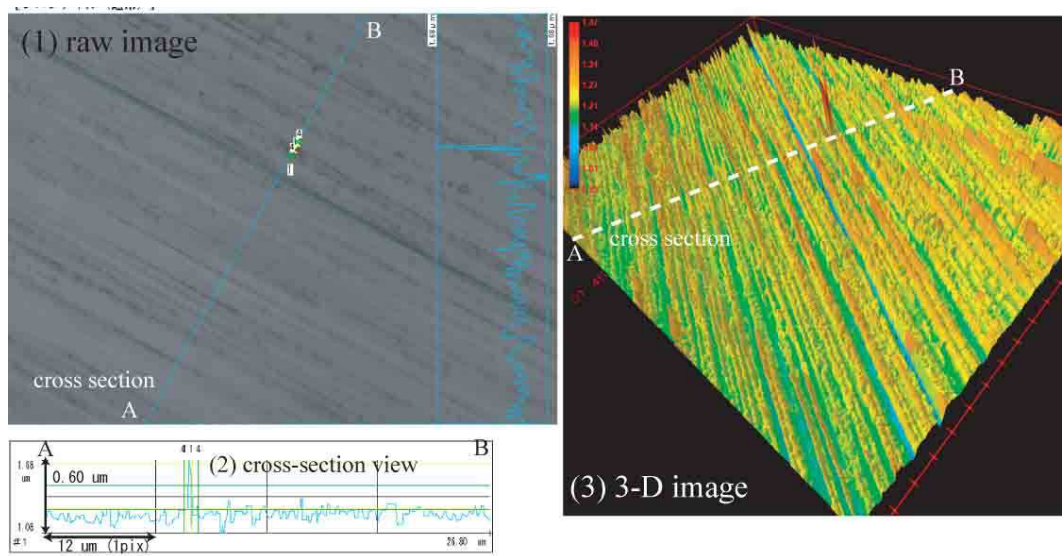}
   \caption[Surface condition of the grinded wafer.]{Surface of
     the ground wafer observed by a color laser three dimensional
     profiling microscope (VK-9500; Keyence). (1)~Raw image of part of
     the ground surface. (2)~Cross sectional view acquired by laser
     ranging. (3)~3-D view.}
  \label{fig:grind_map}
 \end{center}
\end{figure}

To thin the wafers of the tested model of BS-CCD,
we adopted a combination of grinding and polishing methods
\cite{takagi05}. 
We found
that the polishing process often physically destroy the  CCD wafers,
which impacts the yield of the CCD-NeXT1 devices. Thus, we
processed devices with a wafer thickness of $150~{\rm \mu m}$ by omitting
the polishing process in order to examine their feasibility as X-ray
detectors. We examined the surfaces of the devices with a color
laser three-dimensional profiling microscope (VK-9500; Keyence) and
found that numerous microscopic striations were left on the surface due
to the omission of the polishing process.
Figure~\ref{fig:grind_trail} is a close-up view of the ground
surface of the CCD-NeXT1 device, which shows the striations. Figure~\ref{fig:grind_map} shows the
condition of the ground surface. We found the depth and width of the striations to be less than $\sim 0.3~{\rm \mu m}$ peak-to-peak and sub-${\rm
  \mu m}$, respectively. Compared to the wafer thickness ($ \sim
150~{\rm \mu m}$) and the size of pixels ($\sim 12~{\rm \mu m}$), these
trails are negligibly small. Therefore, we concluded that grinding only is
sufficient for the thinning process.

\begin{figure}
 \begin{center}
   \includegraphics[width=4.5cm,angle=270]{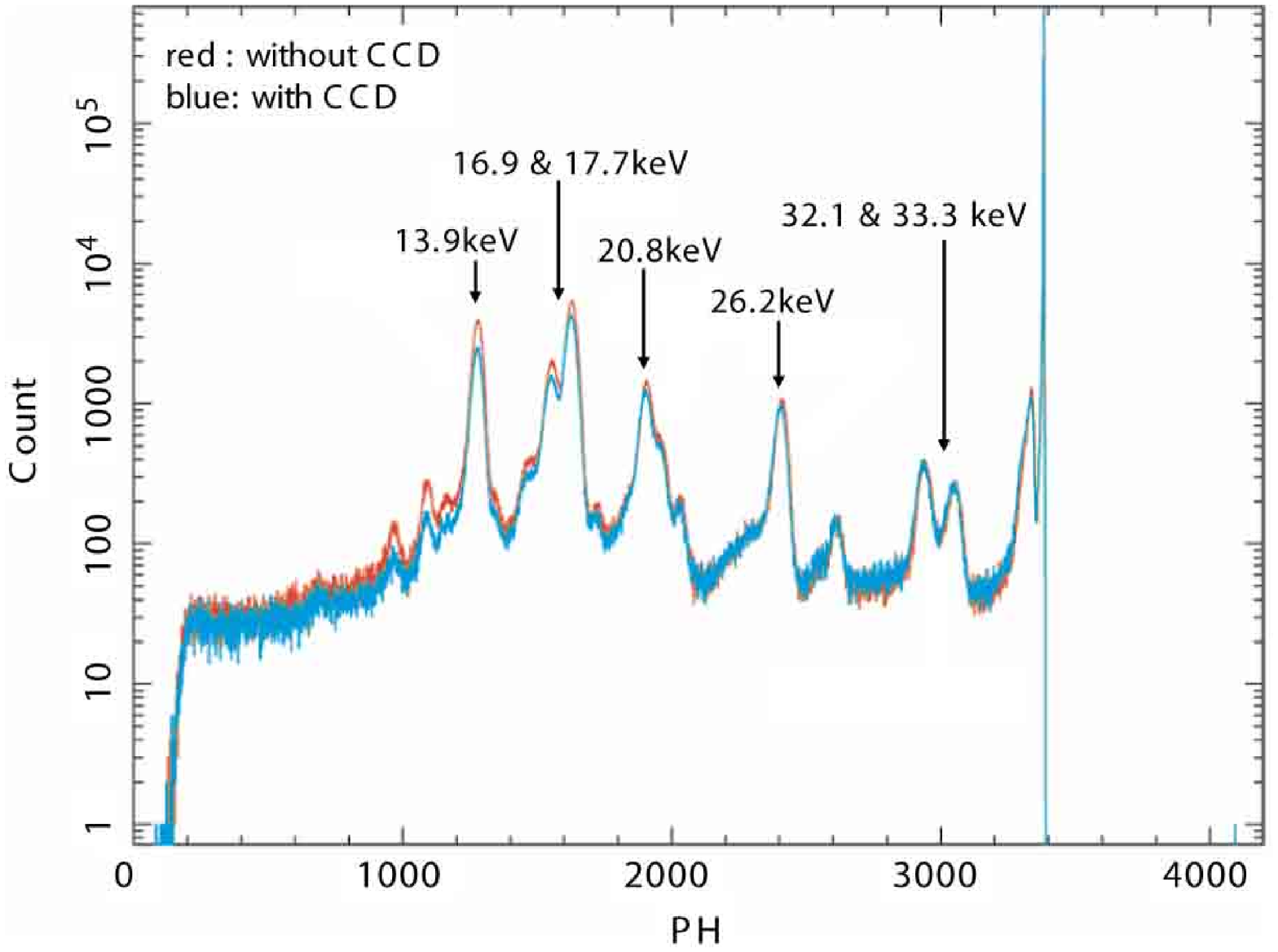}
   \includegraphics[width=4.5cm,angle=270]{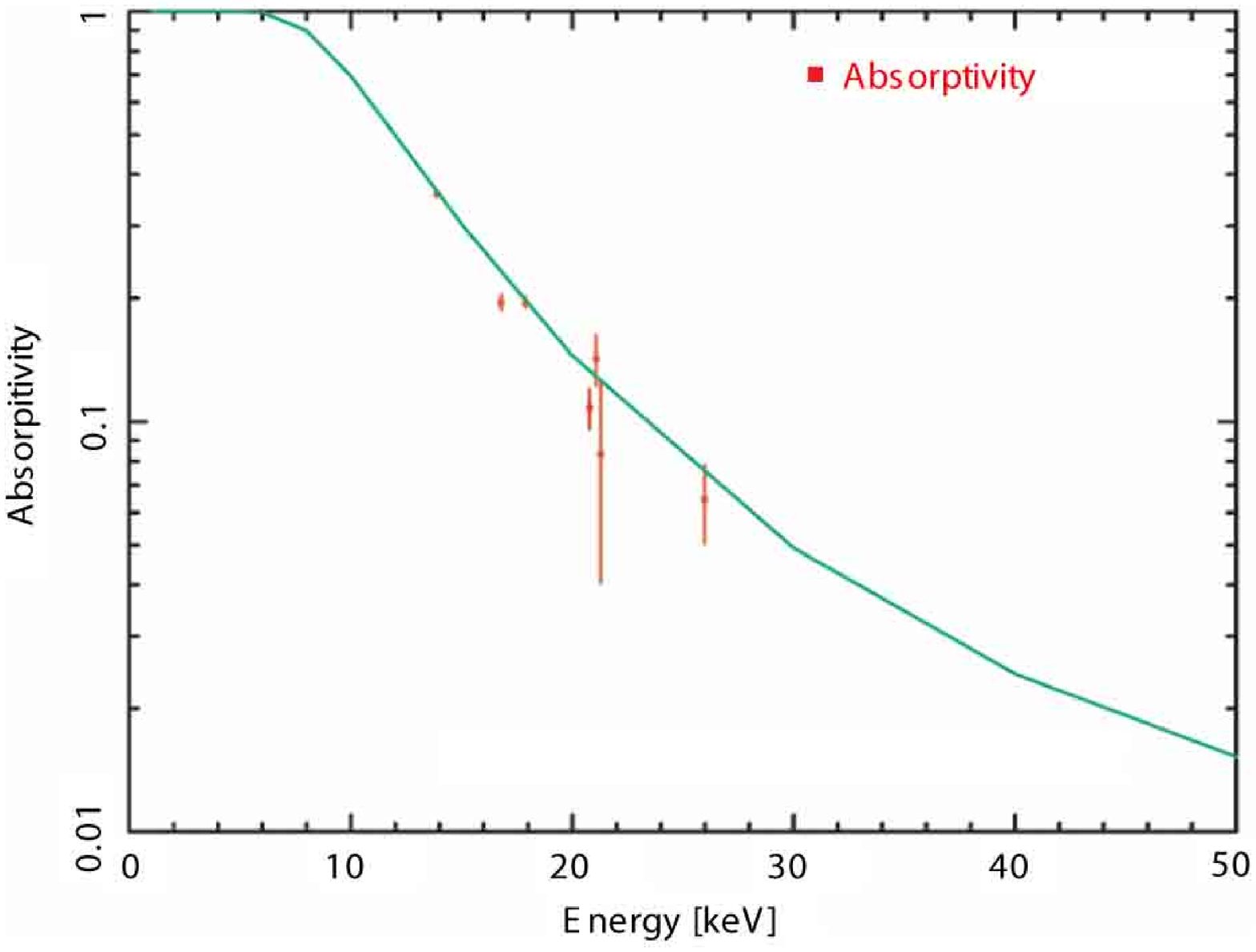}
   \caption[(left) Spectra of $^{241}$Am detected by CdTe detector
     (XR-100T-CdTe; AmpTek). (b) The count ratio at each
     characteristic X-ray line between the absorbed spectrum by the IA
     of CCD-NeXT1 and the unabsorbed one.]  {(left) Spectra of X-rays
     from $^{241}$Am detected by the CdTe detector (XR-100T-CdTe;
     AmpTek). The red line shows the spectrum obtained from the direct 
	exposure of the CdTe detector to the $^{241}$Am source.
	The blue line shows the spectrum absorbed by the
     device. (right) Absorptivity at each characteristic X-ray
     line between the absorbed spectrum and the unabsorbed one.
     Solid curve represents the expected absorptivity for a 
     150-$\rm \mu m$ thick wafer.}  
  \label{fig:NeXT1wafer}
 \end{center}
\end{figure}

We investigated X-ray performance of the thinned devices fabricated
without the polishing process. 
First, we measured the wafer thickness of the device to confirm it was 
$150~{\rm \mu m}$ as designed. We placed the IA (back-supportless
area) of the CCD-NeXT1 device between an  $^{241}$Am source
and the CdTe detector (XR-100T-CdTe; AmpTek) to acquire an
``absorbed'' $^{241}$Am spectrum. Subsequently, we detached the
CCD-NeXT1 device and acquired an ``unabsorbed'' spectrum using the same
experimental setup. 
The left panel in Figure~\ref{fig:NeXT1wafer} shows the two spectra.
The intensity ratio of each
characteristic X-ray line between the absorbed and unabsorbed spectra
shows the X-ray transmission properties ($T_{CCD}$) of this wafer. Therefore the
absorptivity ($A_{CCD}$) was $1-T_{CCD}$. We measured the $A_{CCD}$ at
each characteristic X-ray line and found it is consistent with
a silicon thickness of $150~{\rm \mu m}$ (the right panel of
Figure~\ref{fig:NeXT1wafer}). Thus, we confirmed that the wafer of
the thinned device is 150-${\rm \mu m}$ thick, as designed.

\begin{figure}
 \begin{center}
   \includegraphics[width=6.8cm]{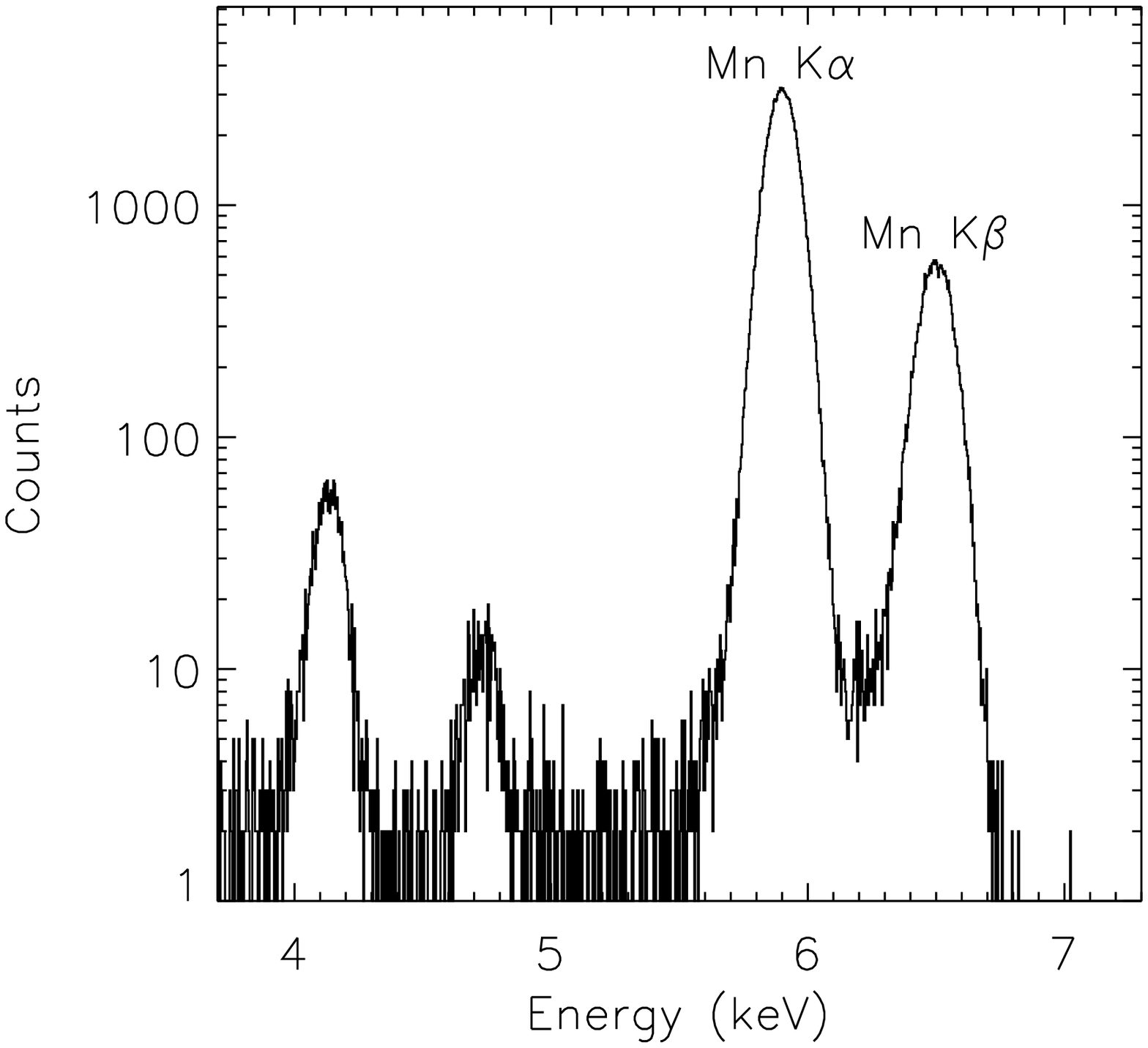}
   \includegraphics[width=6.8cm]{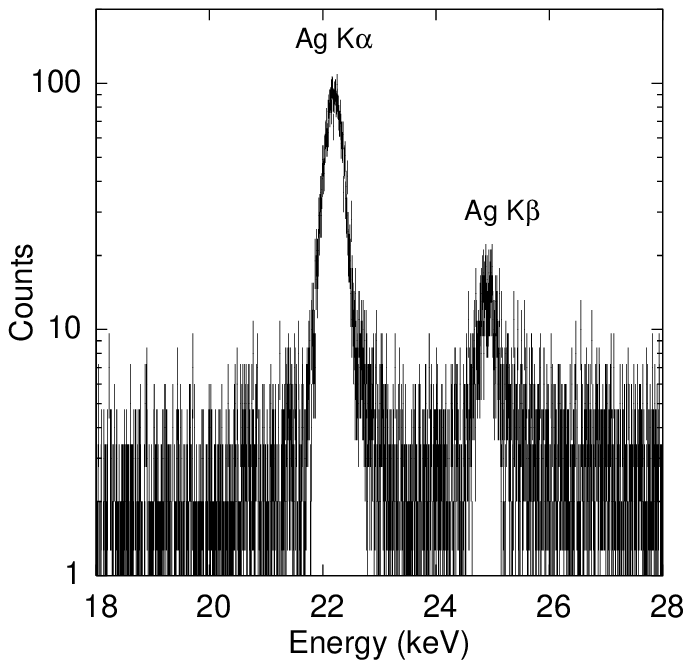}
   \caption[]{Same as those in Figure~\ref{fig:KG5spec}, but obtained
     with the thinned device.}
  \label{fig:KG4spec}
 \end{center}
\end{figure}

Second, we investigated the X-ray performance of the thinned device
by conducting the same experiments mentioned in the previous section for 
an unthinned device. Figure~\ref{fig:KG4spec} shows the
acquired spectra of single pixel events from an $^{55}$Fe source
and {\it ASCA} grade 0--7 events from an $^{109}$Cd source. 
The readout noise and the energy 
resolution were $4.9\pm 0.2 ~{\rm e^-}$ (RMS) and
$135.2\pm 0.2$~eV (FWHM) at the 5.9~keV
(Table~\ref{tab:NeXT1results}), respectively. 
CTIs were $\sim 1 \times 10^{-7}$
both for the vertical and horizontal transfers. 
It should be noted that the CCD was read out through node A.

Finally, we measured the detection efficiency 
by the irradiating the CCD with  the collimated
$^{109}$Cd source as previously described. We obtained a detection efficiency for the two lines
by comparing the obtained count rate with the absolute flux of the
$^{109}$Cd X-rays. In this experiment, the absolute fluxes of
  Ag~K$_\alpha$ and K$_\beta$ were calibrated separately. The
detection efficiencies were 5.5\% and 3.5\% for the Ag~K$_\alpha$ and
K$_\beta$, respectively, implying a 76-${\rm \mu m}$ thick depletion layer.

From the results described above, the X-ray performance of the
thinned device was essentially equivalent to that of the unthinned
device and the test model of BS-CCD~\cite{takagi05}. Hence we concluded that
the thinned CCD-NeXT1 device satisfied the requirements for
the X-ray performance of the baseline plan for the SXI.

\section{Summary}

\begin{itemize}

\item Following the successful development of the test models of
  BS-CCD for SXI onboard the {\it NeXT} satellite, we developed an
  evaluation model, CCD-NeXT1. We fabricated two types of
  CCD-NeXT1. One was a type of un-thinned CCD with
  625-$\mu$m thick wafers. The other type was a thinned CCD with
  150-$\mu \rm m$ thickness.

\item We processed thinned CCD-NeXT1 devices by omitting the polishing
  process in the thinning process. The evaluation of the devices
  confirmed that omitting the polishing process did not 
impact on the X-ray performance. 

\item We did not observe a significant difference in the X-ray performance
  of the unthinned and the thinned devices. The energy resolution
  and the readout noise were $\sim$140~eV (FWHM) at 5.9~keV 
and $\sim$5 electrons (RMS), respectively. 
The detection efficiency of the $\rm ^{109}Cd$ photons indicates 
that the depletion layer is $\sim$ 80-${\rm \mu m}$ thick.
This performance meets the requirements for the baseline plan of the SXI.

\end{itemize}
\section*{ACKNOWLEDGMENTS}

This work is based upon the Grant-in-Aid for the 21st Century COE
``Center for Diversity and Universality in Physics'' from the Ministry of
Education, Culture, Sports, Science and Technology (MEXT) of Japan,
and is supported by a Grant-in-aid (Fiscal Year 2002-2006) for one of
the Priority Research Areas in Japan; ``New Development in Black Hole
Astronomy''. 
This work is partly supported by a Grant-in-Aid for Scientific
Research by
the Ministry of Education, Culture, Sports, Science and Technology
of Japan (16002004).
T.I. is supported by JSPS Research Fellowship for Young
Scientists.


\clearpage

\begin{table}[h]
{\small
 \begin{center}
\caption[Specification of the baseline, goal plans and XIS]
        {Specifications of the baseline plan, goal plan, and XIS.}
  \begin{tabular}{lccc}
   \hline\hline
   Item&Baseline plan &Goal plan&{\it Suzaku} XIS\\
   \hline
   Readout noise                 & $\rm < 5~e^{-}$           & (same)     & $\rm < 3~e^{-}$ \\
   Energy resolution$^{\dagger}$ & $<135~{\rm eV}$          & (same)     & 130~eV  \\
   Wafer type                    & N channel                & P channel  & N channel\\
   Imaging area                  & $50\times 50 ~{\rm mm^2}$ & (same)     & $24.5\times 24.5~{\rm mm^2}$\\ 
   Pixel size$^{\flat}$          & $\rm 24~\times 24~\mu m^{2}$      & (same)     & $\rm 24~\times 24~\mu m^{2}$\\
   Readout nodes                 & 8                   & 8           & 4 \\
   Depletion layer               & 70--100~$\rm \mu m$ & 200~$\rm \mu m$ & $70~{\rm \mu m}$ (FI), $45~{\rm \mu m}$ (BI)\\
   Field-free layer              & 50--80~$\rm \mu m$  & 0~$\rm \mu m$   & $>500{\rm~\mu m}$ (FI), $0{\rm~\mu m}$ (BI) \\
   Illuminated type              & front illuminated  & back illuminated &  both types \\
   Surface coating               & aluminum                  & aluminum  & none \\
   \hline
   \hline
   \multicolumn{3}{l}{\footnotesize $^{\dagger}$: Energy resolution for  Mn~$\rm K_{\alpha}$ (5.9~keV) X-rays (FWHM).}\\
   \multicolumn{3}{l}{\footnotesize $^{\flat}$: Values in imaging area.}\\
   \label{tab:SXIgoala}
  \end{tabular}
 \end{center}
}
\end{table}

\begin{table}[htb]
  \caption[Specification of CCD-NeXT]{Specification of CCD-NeXT1}
  \begin{center}
    \begin{tabular*}{1.0\textwidth}{ll}
      \hline \hline
      Number of pixels & 2000 $\times$ 2000 $\times$2$^{\dagger}$ \\
      Pixel size       & 12 $\mu$m $\times$ 12 $\mu$m  \\
      Clock phase      & 2~phase \\
      Transfer method  & Frame Transfer \\
      Readout nodes    & 2 \\
      Depletion layer  & 70--100~$\rm \mu m$ \\
      wafer thickness  & 150 or 625~$\rm \mu m$ \\
      Illuminated type & front illuminated  \\
      Surface coating  & aluminum            \\
      \hline
      \multicolumn{2}{l}{\footnotesize $^{\dagger}$ Sum of imaging area and frame stored region.}\\
    \end{tabular*}
    \label{tab:NeXT1spec}
  \end{center}
\end{table}

\begin{table}[htb]
  \caption[Performance of CCD-NeXT]{Performance of CCD-NeXT1}
  \begin{center}
    \begin{tabular*}{1.0\textwidth}{lccc}
      \hline \hline
      parameter & \multicolumn{2}{c}{unthinned device} & thinned device\\
      \hline
      readout node                    & A               & B               & A \\
      Readout noise in rms (e$^{-}$) & $4.7\pm 0.2$    & $5.5\pm 0.2$    & $4.9\pm 0.2$  \\
      Energy resolution (eV)$^\dagger$  & $141.8\pm 0.6$  & $146.7\pm 0.8$  & $135.2\pm 0.5$ \\
      Depletion layer (${\rm \mu m}$) & \multicolumn{2}{c}{84}                & 76 \\
      Wafer (${\rm \mu m}$)           & \multicolumn{2}{c}{625 (not thinned)} & 150 \\
      Horizontal CTI $(10^{-7})$   & $0.9\pm 1.4$ & $-6.5\pm 1.4$ & $-1.2\pm 0.5$ \\
      Vertical CTI $(10^{-7})$     & $0.3\pm 0.7$ & $1.5\pm 0.5$  & $-0.3\pm 0.5$ \\
      \hline
      \multicolumn{4}{l}{\footnotesize Errors are at $1\sigma$}\\
      \multicolumn{4}{l}{\footnotesize $^{\dagger}$ Energy resolutions are 
	for Mn~$\rm K_{\alpha}$ (5.9~keV) X-rays (FWHM).}   \\
    \end{tabular*}
    \label{tab:NeXT1results}
  \end{center}
\end{table}

\end{document}